\documentclass[10pt,manuscript]{emulateapj}

\newcommand{\mJy}{\,{\rm mJy} }

\begin{document}

\title{Implications of a High Angular Resolution Image of the Sunyaev-Zel'dovich Effect in RXJ1347-1145}

\author{B.~S. Mason\altaffilmark{1,2}, S.R. Dicker\altaffilmark{3},
  P.M. Korngut\altaffilmark{3}, M.J. Devlin\altaffilmark{3},
  W.D.~Cotton\altaffilmark{1}, P.M. Koch\altaffilmark{4},
  S.M. Molnar\altaffilmark{4}, J. Sievers\altaffilmark{8},
  J.E. Aguirre\altaffilmark{3}, D. Benford\altaffilmark{5},
  J.G. Staguhn\altaffilmark{5,6}, H. Moseley\altaffilmark{5},
  K.D. Irwin\altaffilmark{7}, P.Ade\altaffilmark{8}}

\altaffiltext{1}{National Radio Astronomy Observatory, 520 Edgemont Rd. Charlottesville VA 22903, USA}
\altaffiltext{2}{contact author: bmason@nrao.edu}

\altaffiltext{3}{University of Pennsylvania, 209 S. 33rd St., Philadelphia, PA 19104, USA}

\altaffiltext{4}{Institute of Astronomy and Astrophysics, Academia Sinica.
P.O. Box 23-141, Taipei 10617, Taiwan, R.O.C. }

\altaffiltext{5}{NASA Goddard Space Flight Center, Greenbelt, MD 20771, USA}

\altaffiltext{6}{Johns Hopkins U, Dept. of Physics \& Astronomy, 3400 N. Charles St, Baltimore, MD  21218, USA}

\altaffiltext{7}{National Institute of Standards and Technology, 325 Broadway, Boulder, CO 80303, USA}

\altaffiltext{8}{School of Physics and Astronomy, Cardiff University, 5 The Parade, Cardiff, CF24 3AA, UK}

%
%
%
%
%
%
%

\begin{abstract}

The most X-ray luminous cluster known, RXJ1347-1145 ($z=0.45$), has
been the object of extensive study across the electromagnetic
spectrum.  We have imaged the Sunyaev-Zel'dovich Effect (SZE) at 90
GHz ($\lambda = 3.3$ mm) in RXJ1347-1145 at $10''$ resolution with the
64-pixel MUSTANG bolometer array on the Green Bank Telescope (GBT),
confirming a previously reported strong, localized enhancement of the
SZE $20''$ to the South-East of the center of X-ray emission. This
enhancement of the SZE has been interpreted as shock-heated ($> 20 \,
{\rm keV}$) gas caused by an ongoing major (low mass-ratio) merger
event. Our data support this interpretation. We also detect a
pronounced asymmetry in the projected cluster pressure profile, with
the pressure just east of the cluster core $\sim 1.6 \times$ higher
than just to the west.  This is the highest resolution image of the
SZE made to date.

\end{abstract}


\section{Introduction}

The rich cluster RXJ1347-1145 ($z=0.45$) is the most X-ray luminous
galaxy cluster known \citep{schindler95,schindler97,allen02} and has
been the object of extensive study at radio, millimeter,
submillimeter, optical and X-ray wavelengths
\citep{kitayama04,komatsu01,Gitti07,allen02,schindler97,pointecouteau99,ota08,
  cohen02,bradac08,miranda08}.  Discovered in the ROSAT All-Sky
Survey, RXJ1347-1145 was originally thought to be a dynamically old,
relaxed system \citep{schindler95,schindler97} based on its smooth,
strongly-peaked X-ray morphology--- a prototypical relaxed
``cooling-flow'' cluster. The NOBA 7 bolometer system on the 45-meter
Nobeyama telescope \citep{kitayama04,komatsu01} has made
high-resolution observations ($13''$ FWHM, smoothed to $\sim 19''$ in
the presented map) of the Sunyaev-Zel'dovich effect (SZE) at 150 GHz
which indicate a strong enhancement of the SZ effect $20'' \, (170 \,
{\rm kpc})$ to the south-east of the peak of the X-ray emission,
however. Hints of this asymmetry had been seen in earlier, lower
resolution measurements with the Diabolo $2.1$~mm photometer on the
IRAM 30-m \citep{pointecouteau99}.  The enhancement has been
interpreted as being due to hot ($T_e > 20 \, {\rm keV}$) gas which is
more difficult to detect using X-rays than cooler gas is, owing to the
lower responsivities of imaging X-ray telescopes such as Chandra and
XMM at energies above $\sim 10 \, {\rm keV}$. In contrast, the SZE
intensity is proportional to $T_e$ up to arbitrarily high
temperatures, aside from relativistic corrections which are weak at 90
GHz, so such hot gas stands out.  The feature is consistent with the
presence of a large substructure of gas in the intra-cluster medium
(ICM) shock-heated by a merger, as is seen in the ``Bullet Cluster''
1E0657-56 \citep{markevitch02}; this interpretaion has been supported
by more recent observations \citep[e.g.][]{allen02,ota08}. {\it Thus,
  rather than being an example of a hydrostatic, relaxed system,
  high-resolution SZE observations suggest that the observed
  properties of the ICM in RXJ1347-1145 are strongly affected by an
  ongoing merger.}  This is a striking cautionary tale for ongoing
blind SZE surveys \citep{carlstrom02}, for which useful X-ray data
will be difficult or impossible to obtain for many high-$z$ systems,
as well as a sign that our current understanding of nearby,
well-studied X-ray clusters may be dramatically incomplete.

Reports \citep{komatsu01,pointecouteau01,kitayama04} of a strong
enhancement of the SZE away from the cluster center are based on
relatively low-resolution images compared to the size of the offsets
and features involved.  SZE images at lower frequencies also show show
substantial offsets between the peak of X-ray and SZE emission; for
instance, the 21 GHz \citep{komatsu01} and SZE peak is $\sim 20''$ to
the SE of the X-ray peak, and the 30 GHz \citep{reese02} SZE peak is
$\sim 13''$ to the SE of the X-ray peak.  The situation is further
complicated by the presence of a radio source in the center of the
cluster.  We have sought to test these claims, and to begin to
untangle the astrophysics of this interesting system, with higher
resolution imaging at a complementary frequency.  In this paper we
present the highest angular resolution image of the SZE yet made. We
observed RXJ1347-1145 with the MUSTANG 90 GHz bolometer array on the
Robert C. Byrd Green Bank Telescope (GBT).  At the redshift of the
cluster (and assuming $\Omega_{\Lambda}= 0.3,\Omega_{tot}=1, h=0.73$)
the GBT+MUSTANG $9''$ beam corresponds to a projected length of 54
kpc.  The observations are described in \S~\ref{sec:obs} and the data
reduction in \S~\ref{sec:reduc}. Our interpretation and conclusions
are presented in \S~\ref{sec:concl}.

\section{Instrument \& Observations}
\label{sec:obs}

MUSTANG is a 64 pixel TES bolometer array built for use on the 100-m
GBT \citep{gbtref}. MUSTANG uses reimaging optics with a pixel spacing
of $0.63 f\lambda$, operates in a bandpass of 81--99~GHz, and is
cooled by a pulse tube cooler and Helium-4/Helium-3 closed cycle
refrigerator. Further technical details about MUSTANG can be found in
\citet{dicker2006,dicker2008}. More detailed information is also
provided about MUSTANG, the observing strategy, and the data analysis
algorithms in \citet{agn} and \cite{orion}, which present the results
of other early MUSTANG observations. Further information can be
obtained at the MUSTANG web site\footnote{{\tt
http://www.gb.nrao.edu/mustang/}}.

The observations we present were collected in two runs, one on 21
Feb. 2009 and one on 25 Feb. 2009, each approximately four hours in
duration including time spent setting up the receiver and collecting
calibration observations. For both runs the sky was clear with $\sim 6
\, {\rm mm}$ of precipitable water vapor, corresponding to $\sim 20$~K
zenith atmospheric loading at $3.3 \, {\rm mm}$. Both were night-time
sessions, important because during the day variable gradients in the
telescope structure's temperature degrade its 90 GHz performance
significantly.  The telescope focus was determined by collecting small
maps of a bright calibrator source at a range of focus settings; every
30-40 minutes throughout the session the beam is checked on the
calibrator.  Typically the required focus corrections are stable to a
few millimeters over several hours once residual daytime thermal
gradients have decayed.

Once the focus was established, the in-focus and out-of-focus beam
maps were used to solve for primary aperture wavefront phase errors
using the ``Out-of-Focus'' (OOF) holography technique described by
\citet{bojanoof}. The solutions were applied to the active
surface. This procedure improves the beamshape and increases the
telescope peak forward gain, typically by $\sim 30\%$.  This approach
is effective at correcting phase errors on scales of 20 meters or
larger on the dish surface, but is not sufficiently sensitive to solve
for smaller scales. Therefore there are residual uncorrected wavefront
errors that result in sidelobes out to $\sim 40''$ from the main beam.
Deep beam maps were collected on the brightest 90 GHz sources on
several occasions, principally in test runs on 24/25 March 2009. The
repeatability of the GBT 90 GHz beam after application of the OOF
solutions was found to be good. The analysis of the beam map data is
discussed in \S~\ref{sec:beam}.

Maps of RXJ1347-1145 (J2000 coordinates $13h47m30.5s$,
$-11^{\circ}45'09''$) were collected with a variety of scan patterns
designed to simultanously maximize on-source time and the speed at
which the telescope moved when crossing the source. The effects of
atmospheric and instrumental fluctuations, which become larger on
longer timescales, are reduced by faster scan speeds.  The primary
mapping strategies were: a) a ``daisy'' or ``spirograph'' scan in
which the source of interest is frequently recrossed; and b) a
``billiard ball'' scan, which moves at an approximately constant speed
and has more uniform coverage over a square region of interest. The
nominal region of interest in this case was $5'\times 5'$, centered on
RXJ1347-1145. The size of the maps is sufficiently small that except
under the most exceptionally stable conditions, instrument and random
atmosphere drifts dominate the constant atmosphere (${\rm sec(za)}$)
term and any possible ground pickup. The total integration time on
source was $3.4 \, {\rm h}$.

The asteroid Ceres was observed on both nights and used as the primary
flux calibrator assuming $T_B = 148 \, {\rm K}$ (T. Mueller, private
comm.).  We assign a 15\% uncertainty to this calibration.  We checked
the Ceres calibration on nights when other sources (Saturn, CRL2688)
were visible and found consistent results to within the stated
uncertainty. Using these observations and the lab-measured receiver
optical efficiency of $\eta_{opt,rx} = 50 \pm 10 \%$ we compute an
overall aperture efficiency of $\eta_{aperture,gbt} = 20\%$,
corresponding to a Ruze-equivalent surface RMS of $315 \micron$. This
result is consistent with recent traditional holographic measurements
of the GBT surface. Since the observations presented here the surface
has been set based on further holography maps and now has a surface
RMS, weighted by the MUSTANG illumination pattern, of $\sim 250 \micron$.

\section{Data Reduction}
\label{sec:reduc}

\subsection{Beam Characterization}
\label{sec:beam}

\begin{figure}[h!]
\includegraphics[width = 3.25in, height=2.5in]{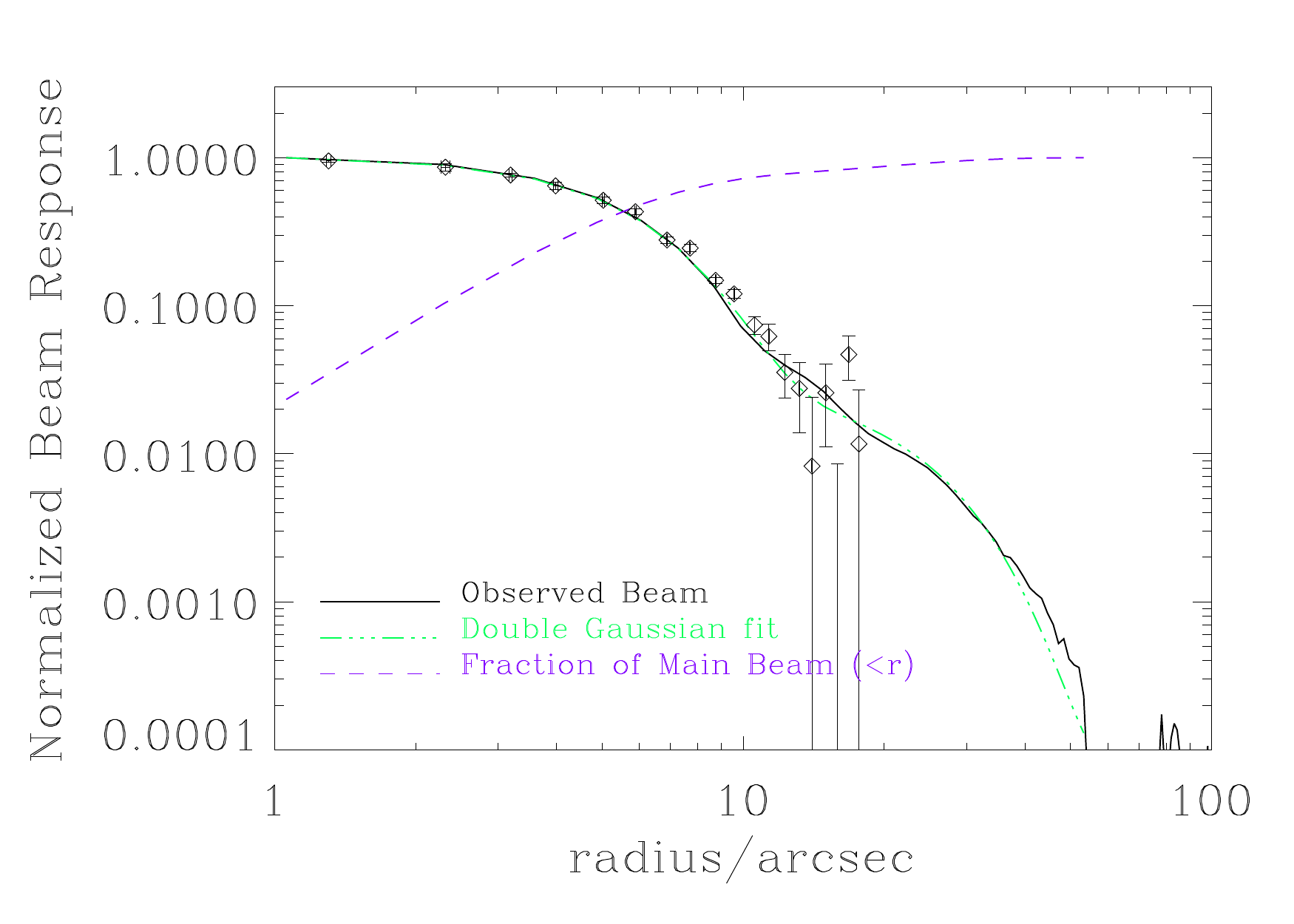}
\caption{MUSTANG+GBT beams determined from observations of 3C279.  The
  green dash triple-dot line shows a double-Gaussian fit to the
  observed beam. The purple dashed line is the cumulative fractional
  beam volume.  We attribute the excess power in the wings of the
  observed beam to residual medium and small scale phase errors on the
  dish. The data points (diamonds with error bars) show a
  complementary determination of the beam from the 5 mJy radio source
  in the center of RXJ1347-1145. The beam in this case is slightly wider
  due to the smoothing (4'' FWHM) applied to the final map, which is accounted
  for in the analysis.}
\label{fig:beams}
\end{figure}

Imaging diffuse, extended structure requires a good understanding of
the instrument and telescope beam response on the sky. To achieve this
we collected numerous beam maps through our observing runs, including
several deep beammaps on bright ($5 \, {\rm Jy}$ or more) sources.
After applying the Out-of-Focus holography corrections to the aperture
the beam results were repeatable; Figure~\ref{fig:beams} shows the
radial beam profile from maps of a bright source (3C279) collected on
two occasions. We find a significant error beam concentrated around
the main lobe which increases the beam volume from $87 \, {\rm
  arcsec^2}$ (for the core component only) to $145 \, {\rm
  arcsec^2}$. We attribute this beam to residual medium and small
scale phase errors on the primary aperture.  The beam shape and volume
is taken into account when comparing to model predictions. By way of
comparison, Figure~\ref{fig:beams} also shows the profile of the beam
determined from the radio source in the center of RXJ1347-1145.
Since the SZ map has been smoothed, the apparent beam is slightly
broader, but allowing for this, still consistent with the beam
determined on 3C279.


\subsection{Imaging}

A number of systematic effects must be taken into account in the time
domain data before forming the image:
\begin{enumerate}

\item The responsivities of individual detectors are measured using an
  internal calibration lamp that is pulsed periodically. Optically
  non-responsive detectors ($10$-$15$ out of 64) are flagged for removal
  from subsequent analysis. Typical detector responsivities are stable
  to $2-3\%$ over the course of several hours.

\item Common mode systematic signals are subtracted from the
  data. These are caused by atmospheric and instrumental (thermal)
  fluctuations.  The pulse tube cooler, which provides the 3K base
  temperature of the receiver, induces a $1.4$~Hz signal due to small
  emission fluctuations of the 3K optics. The pulse-tube signal is
  removed by fitting and subtracting a $1.41$ Hz sine wave. The
  remaining common mode signal is represented by a template formed by
  a weighted average of data from good pixels; this template is
  low-pass filtered and subtracted from the data, with a fitted
  amplitude per detector. The low-pass filter time constant (typically
  $0.1 \, {\rm Hz}$) is determined by the stability of the data in
  question. This procedure helps to preserve large-scale structure in
  the maps.

\item Slow residual per-pixel drifts are removed using low-order orthogonal
  polynomials.

\item Individual detector weights are computed from the residual
  detector timestreams after the above steps.  Since the noise level
  of the detectors varies considerably this is an important step. Best
  results are obtained by retaining only the top $\sim 80\%$ of
  responsive detectors.

\item The remaining calibrated detector timestreams are inspected
  visually on a per-scan (typically 5 minute period) basis. Scans
  which have timestreams with obvious, poorly-removed systematic
  signals remaining are removed. This results in flagging $28\%$ of
  scans. The SNR in an individual detector timestream is sufficiently
  low that this does not bias our map.

\end{enumerate}
Following these calibration steps the detector timestream data are
gridded onto a $2''$ pixellization in Right Ascencion and Declination
using a cloud-in-cell gridding kernel.

To check our results we have implemented three, mostly independent
analysis pipelines. The results in this paper are based on a
straightforward but flexible single-pass pipeline written in IDL,
described above.  There is also an iterative, single-dish CLEAN based
approach implemented in the OBIT package \citep{obit} and an optimal
SNR method in which the time domain data are decomposed into noise
(covariance) eigenvectors; their temporal power spectra computed; and
a maximum likelihood map constructed from the noise-weighted
eigenvectors. Results obtained with these algorithms were consistent.
The first two approaches are described in more detail in
\citet{orion}.

\begin{figure*}[h!]
\includegraphics[width=6in]{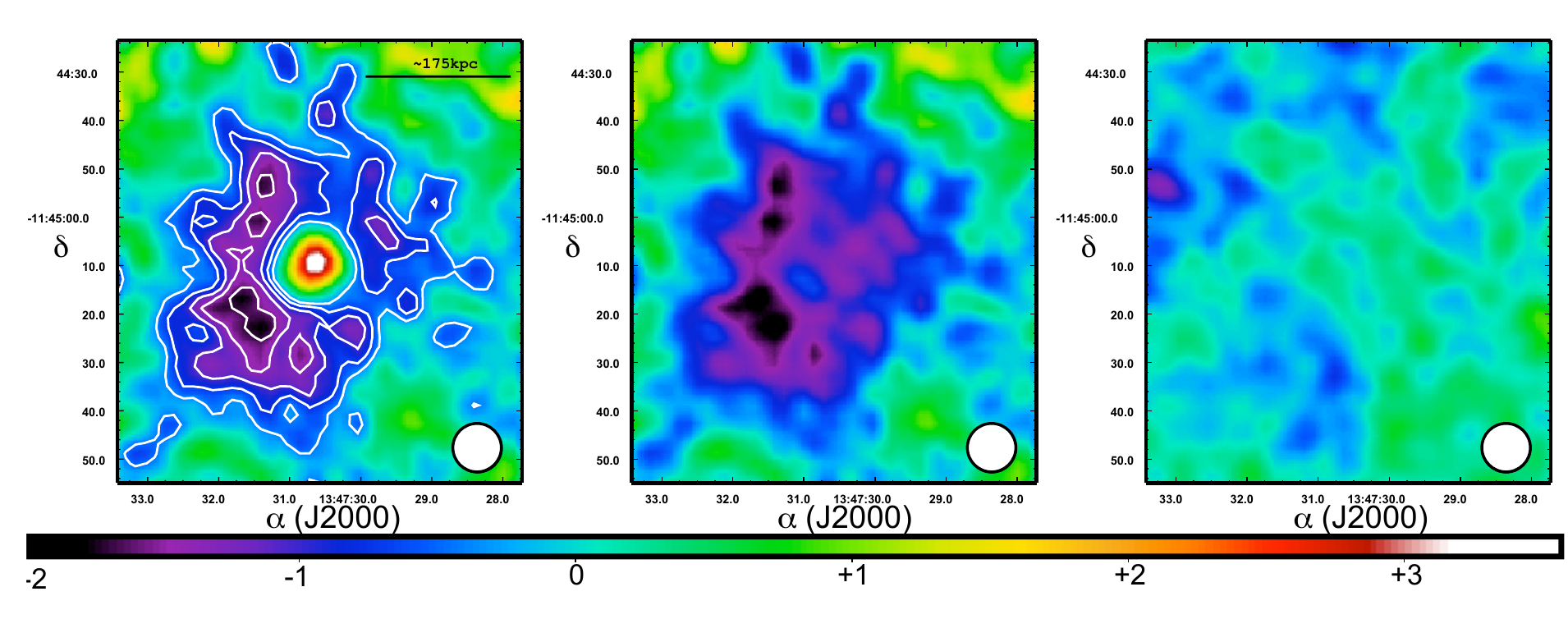}
\caption{MUSTANG image of the SZE in RXJ1347-1145 (left); the same,
  with the point source subtracted as described in the text (center);
  and the individual nights imaged separately and differenced, on the
  same color scale (right). The noise in the center of the map is is
  $\sim 0.3 \, {\rm mJy/bm}$; contours in the left panel correspond to
  SNR of 1 to 5 in $1\sigma$ increments and account for variations in
  integration time in the map, so are not directly proportional to the
  image. Color scale units are mJy/bm. The MUSTANG beam ($10''$ FWHM
  after smoothing) is shown in the lower right of each panel.}
\label{fig:finalmap}
\end{figure*}

Our final map, smoothed by a $4''$ FWHM Gaussian and gridded on
$0''.5$ pixels, is shown in 
Figure~\ref{fig:finalmap}, along with the difference between the two
individual night maps. It shows a strong, clear SZ decrement, well
separated from the central point source and consistent with the level
expected from the \citet[][hereafter K04]{kitayama04} 150 GHz
measurement.  The right hand panel shows the image formed by
differencing the images of the two individual nights. By computing the
RMS in a fiducial region in the center of the difference image (and
scaling down by a factor of $2$ to account for the differencing and
the shorter integration times) we estimate a map-center image noise of
$\sim 0.3 \, {\rm mJy/bm}$ (rms). The noise level in regions of the
map outside the fiducial region is corrected for exposure time
variations assuming Gaussian, random noise with a white power
spectrum.  The enhancement of the SZE to the south east of the X-ray
peak, originally detected by Komatsu et al. at $4.2\sigma$
significance, is confirmed by our measurement at $5.4 \sigma$
(indicating the peak SNR per beam) with a factor of $\sim$ 2 greater
angular resolution.

A detailed assessment of the impact of this is presented in
\S~\ref{sec:sims}. Work is underway to develop analysis techniques
which account for correlated noise in a way that permits quantitative
model fitting.

\subsection{Simulations}
\label{sec:sims}

It is difficult to measure diffuse, extended structure such as the
SZE, particularly in the presence of potentially contaminating
systematic signals such as time-varying atmospheric fluctuations.  To
assess the impact of residual, unmodelled noise fluctuations in the
maps we have undertaken an extensive suite of simulations which
replace the raw detector timestream data with simulated data. As a
source for the simulated data we used real detector timestreams
collected during observations of a blank patch of sky collected for
another project. The phase of these timestreams with respect to the
telescope trajectory on RXJ1347-1145 was randomly shifted to create
different instances of noisy cluster observations.  We added simulated
astronomical signals as described below in order to determine how well
the (known) input signals are recovered in the maps.

To assess the spatial fidelity of our reconstructed images, random
white-noise skies were generated on $2''$ grids, subsequently smoothed
by a $9''$ (FWHM) Gaussian.  These skymaps served as input to generate
fake timestreams which were then processed by the exact processing
scripts used to produce the image in Figure~\ref{fig:finalmap}.  The
ratio of the absolute magnitude of the Fourier transform of the
reconstructed sky map to the absolute magnitude of the Fourier
transform of the input skymap measures the fidelity of our image
reconstructions as a function of angular scale. The results of
repeating this 100 times, with different white-noise skies and noise
instances, are shown in Figure~\ref{fig:sim}. We find that our
pipeline faithfully recovers structures up to $60''$, with reasonable
response but some loss of amplitude on larger scales, up to
$120''$. The loss of structure on small angular scales is an effect of
our relatively coarse pixellization.  Simulations were carried out at
similar signal to noise ratios as those in our final map, although
changes in the signal to noise ratio of over a factor of 5 showed no
significant change to our transfer function

The common-mode subtraction, essential to removing atmospheric and
instrumental systematic signals, can also introduce negative bowls
around bright point sources which could mimic the SZE in cases such as
RXJ1347-1145.  To determine the magnitude of this systematic we have
followed a similar approach. Instead of white-noise skies the input
signal consists of a single unresolved source with a flux density of
5~mJy at the location of the radio source seen in RXJ1347-1145.  The
resulting negative bowl in the reconstructed images has a mean peak
spurious decrement $\sim 2\%$ of the point source peak brightness, in
comparison with $\sim 50\%$ for our real data.  Additionally the
iterative pipeline (OBIT) is much less susceptible to such artifacts,
and shows consistent results. We conclude that this is not a
significant contribution to our result.

\subsection{Image Domain Noise Estimate}

\begin{figure}[h!]
\includegraphics[width = 3.25in, height=2.5in]{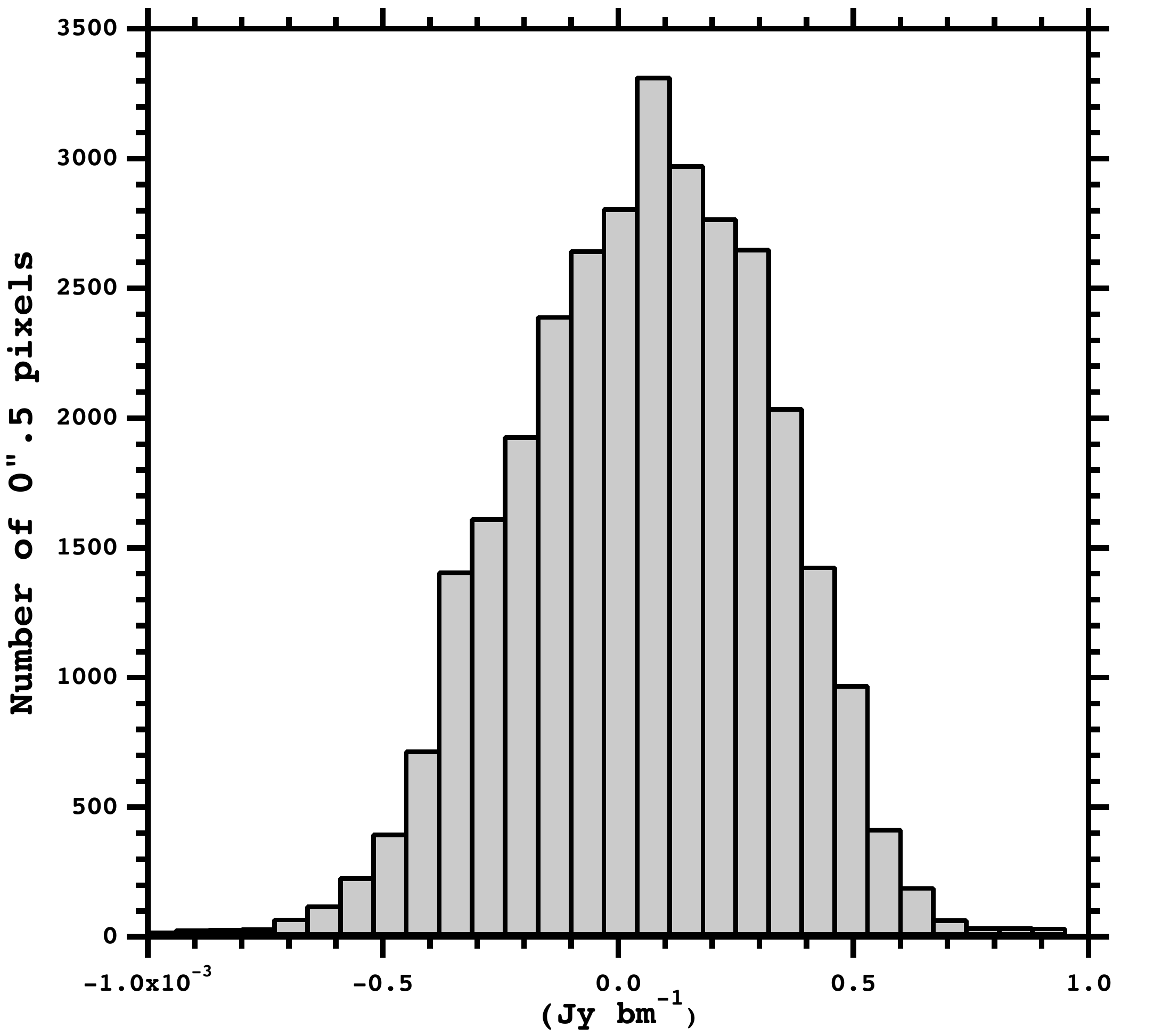}
\caption{Histogram of pixel values in the difference image in
  Figure~\ref{fig:finalmap}. The histogram is well described by a
  gaussian distribution of  $1 \sigma = 0.3 \, {\rm mJy/bm}$}
\label{fig:hist}
\end{figure}

We divide the data set in half and subtract the individual night
images to obtain a difference map. The RMS of this difference map in
the central 85 by 93 arcseconds, dividing by two to correct for the
differencing and the reduced integration time in each individual night
image, gives an image noise level of $0.3 \, {\rm mJy}$. A histogram
of the pixel values in this region of the difference image is shown in
figure \ref{fig:hist}.

We obtain a complementary estimate of the noise from a region of the
final SZ map well away from the cluster. Correcting for the
difference in integration times in these regions of the map, this
result is consistent to within 8\%.

\begin{figure}[h!]
\epsscale{1.15}
\includegraphics[width = 3.5in, height=3in]{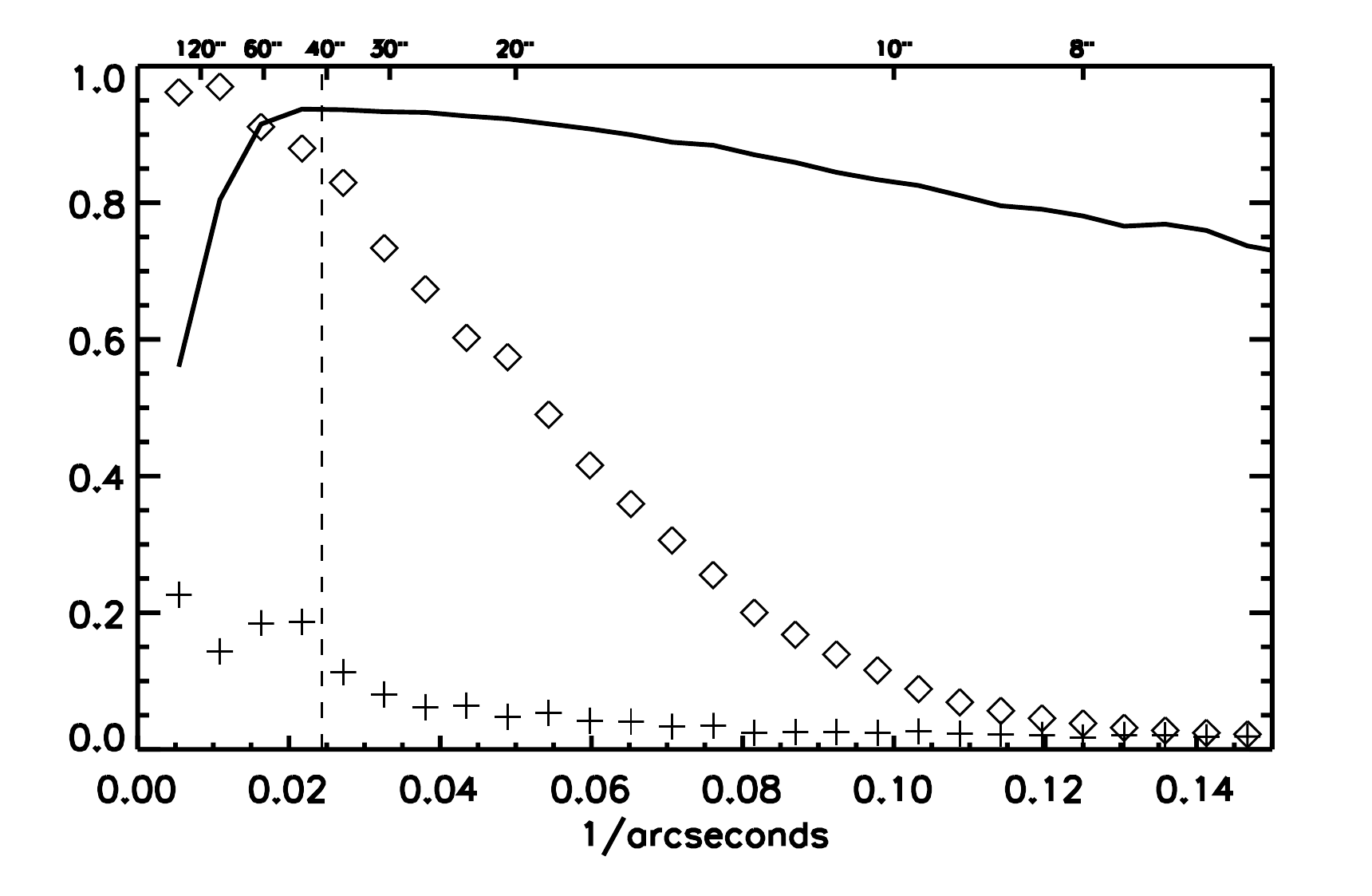}
\caption{Map noise characteristics. {\it Solid curve}: The ratio of
  the absolute magnitude of the Fourier transform of the input map to
  the absolute magnitude of the Fourier transform of the output map
  from simulations.  All structure on scales smaller than $1'$ was
  recovered well although there is a fall off towards high spatial
  frequencies due to pixellization effects.  {\it Diamonds}: The
  absolute magnitude of the Fourier transform of the (reconstructed
  simulated) cluster map, normalized to a peak value of unity.  {\it
    Plus marks}: Absolute magnitude of the fourier transform of
  signal-free simulated maps, with the same normalization as the
  cluster data.  The dashed line shows the size of the MUSTANG
  instantaneous field of view.}
\label{fig:sim}
\end{figure}

Using this noise figure, the peak SNR per beam in the map--- on the SZE
decrement SE of the cluster core--- is $5.4\sigma$. More aggressive
filtering the map results in an even higher detection significance for
the SE enhancement by reducing the low spatial frequency tail of
the noise power spectrum (Figure~\ref{fig:sim}). 


\subsection{The Effects, and Subtraction, of the Central Radio Source}

Our final map has sufficient angular resolution to distinguish the
central radio source from the structures of interest. In particular it
is clear that, as seen in earlier analyses of the SZE in this cluster
\citep{pointecouteau99} there is a strong azimuthal variation in the
intensity of the SZE at a radius of $\sim 20''$ from the X-ray
centroid, which also coincides with the radio source.

To produce a source-subtracted image we fit and subtract an
azimuthally symmetric, double-Gaussian beam (as determined from 3C279
in \S~\ref{sec:beam}). The reason for assuming azimuthal symmetry is
that the hour angle sampling of the 3C279 data is considerably more
limited than that of the RXJ1347-1145 data; therefore the 3C279 data
will not provide a good measurement of the effective two-dimensional
beam, only of its average radial profile. Furthermore the SNR on the
point source in RXJ1347-1145 is insufficient to measure significant
departures from azimuthal symmetry.

The average radial profile of the central source in RXJ1347-1145 is shown in
Figure~\ref{fig:beams} out to $r=15''$, where the signal becomes too
weak to measure above thermal noise and variations in the SZE.


\subsection{Effect of Background Anisotropies}

The angular scales reconstructed in our map ($\sim 1'.5$ and smaller)
correspond to spherical harmonic multipoles of $\ell = 7200$ and
higher. On these scales intrinsic CMB anisotropies are strongly
suppressed by photon diffusive damping at the last scattering surface
and do not contribute measurably to our result at the sensitivity
level we have achieved.

\begin{figure}[h!]
\includegraphics[width = 3.5in, height=3.5in]{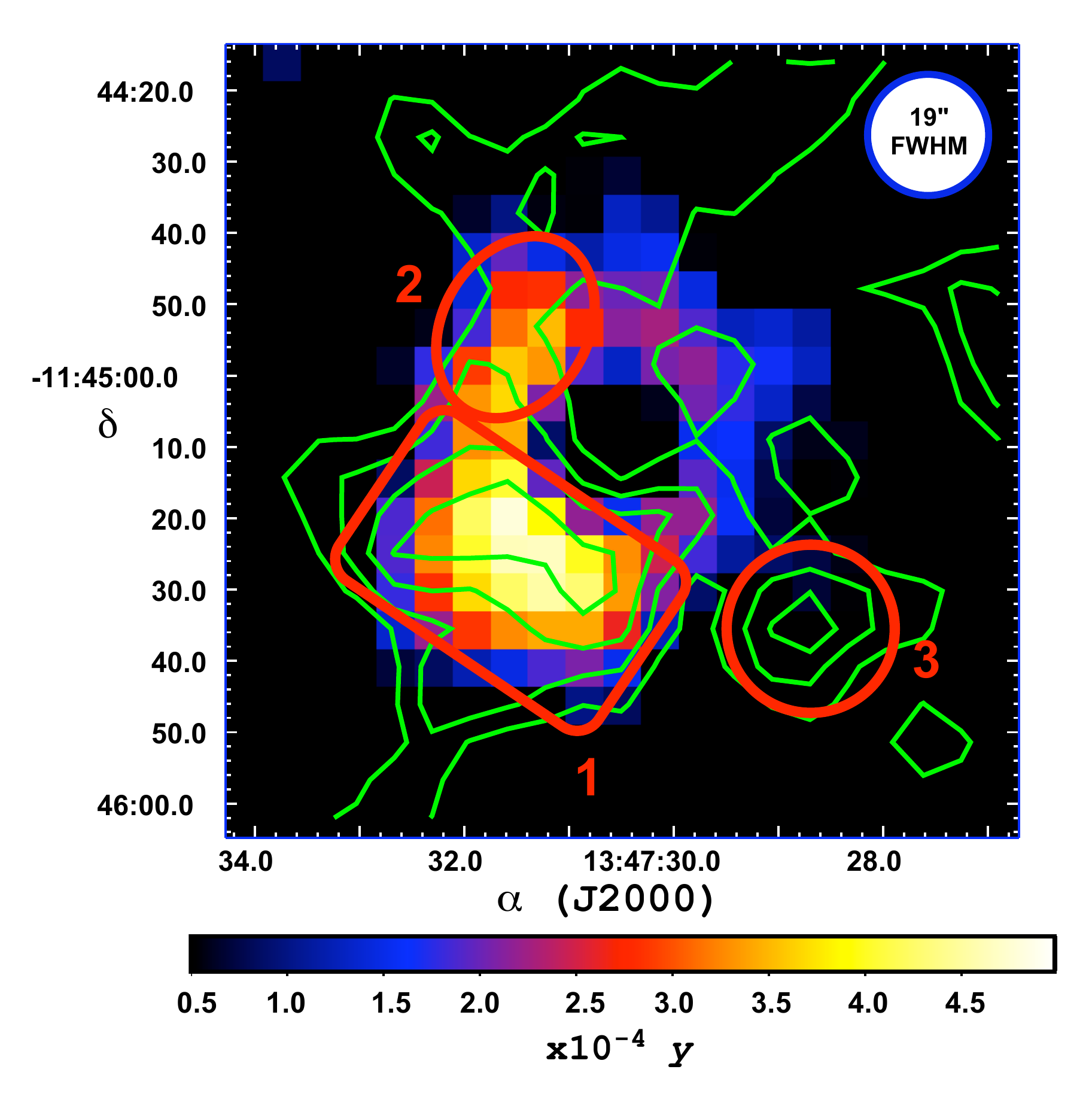}
\caption{Comparison of NOBA and Mustang maps of RX J1347-1145. The
  color scale shows the MUSTANG data with $5''$ pixels, smoothed to
  match the published NOBA map resolution. The contours show the NOBA
  map at intervals of $1.16\times 10^{-4}$ in $y$ starting
  from $5\times 10^{-5}$.  Three labeled features are discussed in the
  text: {\it 1}, the hot shock south-east of the cluster core; {\it 2}, an
  enhancement in integrated pressure to the east; and {\it 3}, a
  compact decrement observed at $\sim 3 \sigma$ by NOBA that is absent
  from the MUSTANG image.}
\label{fig:Noba}
\end{figure}

\begin{figure}[h!]
\includegraphics[width = 3.25in, height=3in]{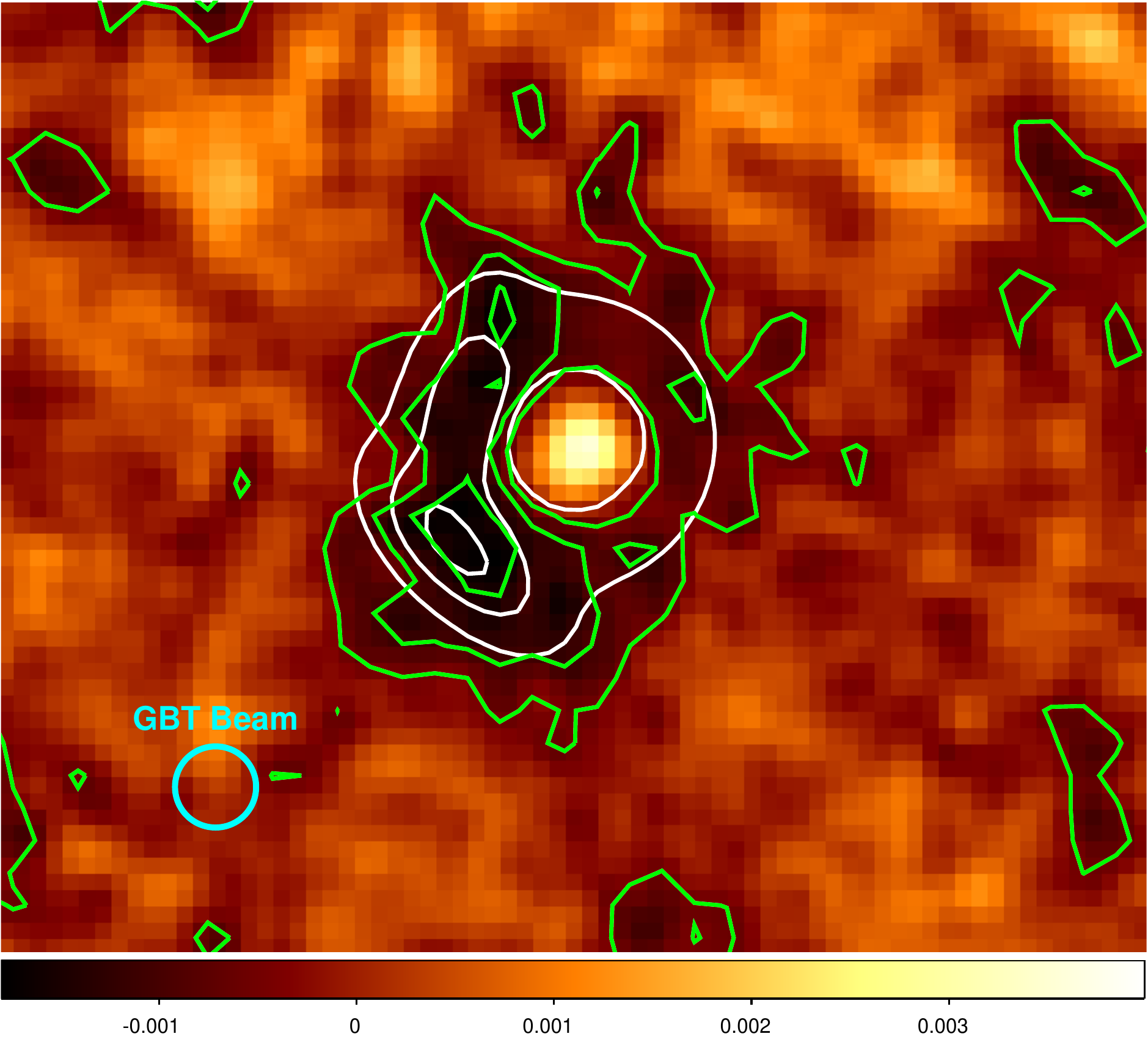}
\caption{MUSTANG SZE image of RXJ1347-1145 with contours (thin green
lines) at  $-1.5$, $-1.0$, and $-0.5$ mJy/bm. The bold white contours,
at the same surface brightness levels, show the model SZ signal
discussed in the text.}
\label{fig:szmodel}
\end{figure}

\begin{figure*}
\includegraphics[width=6in]{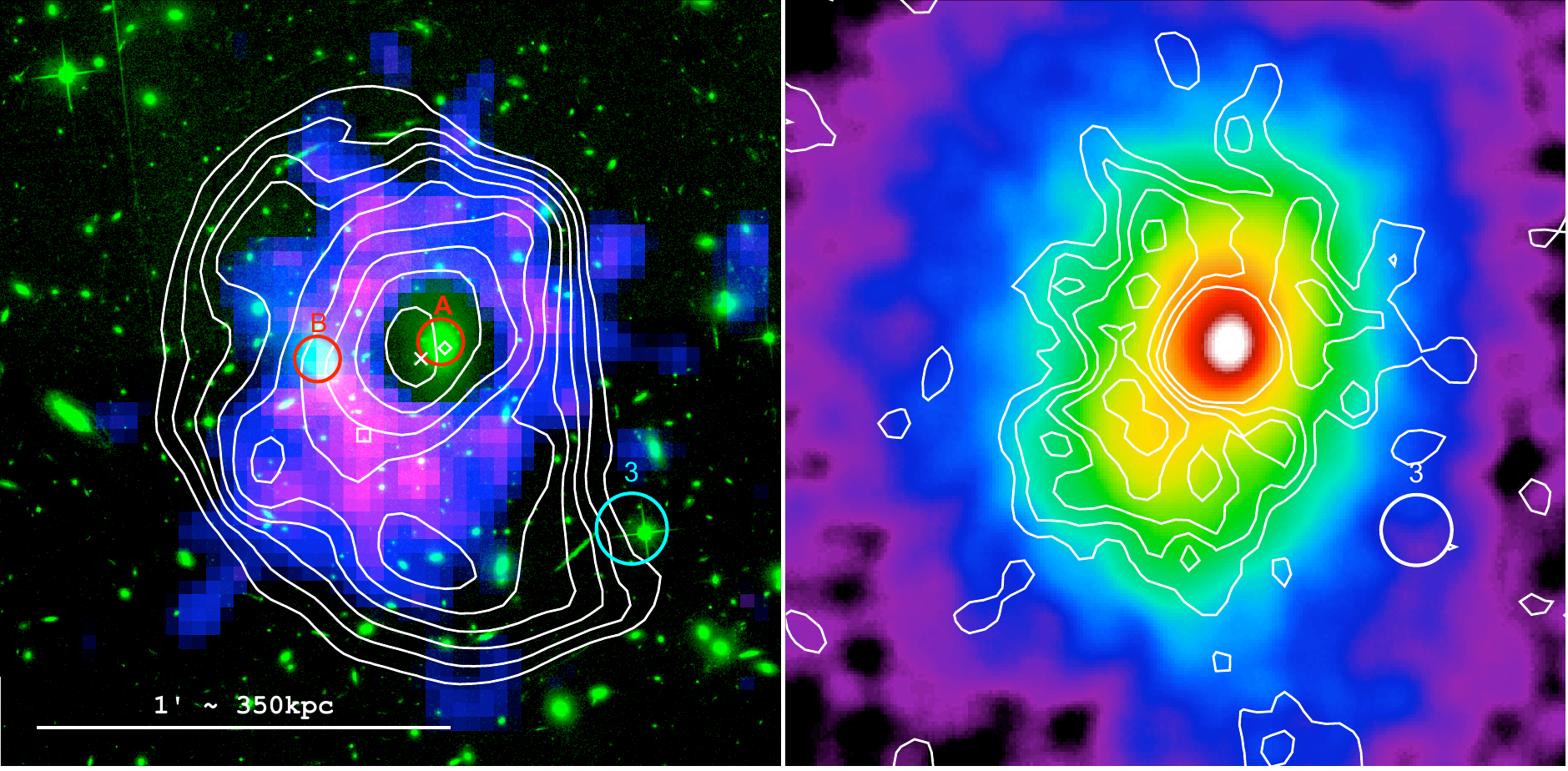}
\caption{{\bf Left:} False color composite Image of
  RXJ1347-1145. Red/blue: Mustang SZ. Green: Archival HST/ACS image
  taken through the F814W filter; and white contours: Surface mass
  density $\kappa$ from the weak + strong lensing analysis of \cite{bradac08}.
  Contours are linearly spaced in intervals of $\Delta\kappa = 0.1$
  beginning at $\kappa = 1.0$. Several features are labelled: {\it A}
  indicates the central BCG, which is a radio source; {\it B}
  indicates the BCG of the secondary cluster; and {\it 3} (also
  labeled in the right-hand panel) indicates the location of the
  discrepancy between NOBA and MUSTANG, discussed in the text and
  Figure~\ref{fig:Noba}.  The diamond, cross and box mark the
  locations of the peaks in X-ray surface brightness, surface mass
  density and SZE decrement respectively.  {\bf Right:} Contours of
  the MUSTANG decrement SNR ($1\sigma$ to $5 \sigma$ in $1\sigma$
  increments) superposed on the Chandra count-rate image smoothed to
  $10''$ resolution.}
\label{fig:composite}
\end{figure*}



\section{Interpretation \& Conclusions}
\label{sec:concl}

\subsection{Comparison with Previous SZE Observations}
Figure~\ref{fig:Noba} presents a direct comparison of the MUSTANG and
NOBA results in units of main-beam averaged Compton y parameter. For a
more accurate comparison, we downgrade the resolution and pixelscale
of the MUSTANG map to match that of NOBA (13'' FWHM on a 5'' pixel
grid). The overall agreement between the maps is excellent, in particular as
regards the amplitude and morphology of the local enhancement of the
SZE south-east of the cluster core.  The largest discrepancy is south
west of the cluster, where NOBA shows a $3\sigma$ compact decrement
which is absent from the MUSTANG data.  Considering the low and
uniform X-ray surface brightness in the vicinity of this discrepancy
(see Figure~\ref{fig:composite}) and the higher angular resolution and
lower noise of the MUSTANG data, it is likely that this feature is a
spurious artifact in the NOBA map.  Both datasets also show a ridge
extending north from the shock front on the eastern side of the
cluster. In the 150 GHz map the feature is of marginal significance
($1-2\sigma$); interestingly, it is clearly visible in the 350 GHz SZE
increment map but K04 dismiss it due to the possibility of confusing
dust emission from the nearby galaxies.  

\subsection{Empirical Model of the SZE in RXJ1347-1145}
We construct a simple empirical model for the cluster SZE assuming the
isothermal $\beta$-model of \citet{schindler97} normalized by the SZE
measurement of \citet{reese02} and \citet{kitayama04} to describe the
bulk cluster emission. We add a 5 mJy point source in the cluster
core, coincident with the peak of the $\beta$-model, and two Gaussian
components in integrated pressure, one south-east and one almost
directly east of the cluster center. In comparing to our 90 GHz data,
we use the relatavistic correction of \citet{sazonov98}, assuming $kT = 25 \,
{\rm keV}$ (which reduces the amplitude of the decrement by 15\%) for
the Gaussian components and $kT = 10 \, {\rm keV}$ for the bulk
component.  The parameters chosen (two Gaussian widths for each
component, a position, a peak surface brightness, and a position
angle) are shown in Table~\ref{tbl:szmodel}.  The resulting sky image
is convolved with our PSF (\S~\ref{sec:beam}) and transfer function
(\S~\ref{sec:sims}).  We find that this provides a good match to the
data (Figure~\ref{fig:szmodel}). The peak comptonization at $10''$
Gaussian resolution is $3.9 \times 10^{-4}$ on the eastern ridge and
$6.0 \times 10^{-4}$ on the region identified as a shock by Komatsu et
al. When convolved to $19''$ FWHM (NOBA) resolution, we find $\Delta y
= 3.9 \times 10^{-4}$, close to their observed value $\Delta y = 4.1
\times 10^{-4}$. The intent of this static, phenomenological model is
simply to provide a description of the observed high
angular-resolution SZE and a direct comparison of NOBA and MUSTANG
results. Work is underway which will allow quantitatively determining
the best fit physical model by simulataneously fitting datasets at
multiple wavelengths using a Monte-Carlo Markov Chain.  This work is
beyond the scope of this paper and will be presented in a follow up
publication.
\begin{table*}
\begin{center}
\begin{tabular}{llll}
Component & Amplitude & Offset  & Notes \\ 
          &      [$y/10^{-3}$]                    &     [$''$]          &       \\ \hline
$\beta$-model & $1.0$ & $0,0$ & $\theta_c = 10''$, $\beta=0.60$ \\
Shock & $1.6$ & -14, 14 & $\sigma_1=8''$, $\sigma_2=2''$, ${\rm P.A.}=45^{\circ}$ \\
Ridge & $1.0$ & 10, 14 & $\sigma_1=8''$, $\sigma_2=2''$, ${\rm P.A.}=-15^{\circ}$ \\ \hline
\end{tabular}
\caption{Note: Offset is  (north,east) of peak X-ray position}
\label{tbl:szmodel}
\end{center}
\end{table*}


\subsection{Multi-wavelength Phenomenology}

Our data show an SZE decrement with an overall significance of $5.4
\sigma$.  At the center of the cluster, coincident with the peak of
X-ray emission and the brightest cluster galaxy (BCG), there is an
unresolved $5 \mJy$ radio source. This flux density is consistent with
the 90 GHz flux density presented in \citet{pointecouteau01}, as well
as what is expected from a power law extrapolation of $1.4$ GHz and 30
GHz measurements \citep{NVSS,reese02}.  A strong, localized SZE
decrement can be seen $20''$ to the south-east of the center of X-ray
emission and clearly separated from the cluster center. Our data also
indicate a high-pressure ridge immediately to the east of the cluster
center.

K04 tentatively attribute the south-east enhancement to a substructure of gas
$240 \pm 183 \, {\rm kpc}$ in length along the line of sight, at a
density (assumed uniform) of $(1.4 \pm 0.59) \times 10^{-2} \, {\rm
  cm^{-3}}$ and with a temperature $T_e = 28.5 \pm 7.3 \, {\rm keV}$.
Recent X-ray spectral measurements \citep{ota08} with SUZAKU also
indicate the presence of hot gas in the south-east region ($T_e =
25.1^{+6.1}_{-4.5} \ ^{+6.9}_{-9.5} \, {\rm keV}$ with statistical and
systematic errors, respectively, at 90\% confidence
level). \citet{allen02} have reported that the slight enhancement of
softer X-ray emission in this region seen by Chandra is consistent
with the presence of a small substructure of hot, shocked gas.
\citet{kitayama04} attribute the hot gas to an ongoing merger in the
plane of the sky. The merger hypothesis is supported by optical data,
in particular, the presence of a second massive elliptical $\sim 20''$
directly to the east of the BCG that coincides with the center of
X-ray emission (and with the radio point source).  Furthermore the
density and temperature of the hot substructure indicate that it is
substantially overpressured compared to the surrounding ICM. Assuming
a sound speed of $1600 \, {\rm km/sec}$ this overpressure region
should relax into the surrounding ICM on a timescale $\sim 0.1 \, {\rm
  Gyr}$, again arguing for an ongoing merger.

Our data support this merger scenario.  To put them in context,
Figure~\ref{fig:composite} shows a composite image with archival
Chandra and HST data, and the weak + strong lensing mass map of
\citet{bradac08}.  We propose that the data are best explained by a
merger occuring in or near the plane of the plane of the sky.  The
left-hand (``B'') cluster, having fallen in from the south-{\it west},
has just passed closest approach and is hooking around to the north-west.
As the clusters merge shock forms, heating the gas in the wake of its
passage. As argued by \cite{kitayama04}, and seen in simulations
\citep{takizawa99}, the clusters must have masses within a factor of 2
or 3 of equality and a substantial ($\sim 4000 \, {\rm km/sec}$)
relative velocity in order to produce the high observed plasma
temperatures, $T_e > 20 \, {\rm keV}$.  This merger geometry is
consistent with the lack of structure in the line-of-sight cluster
member galaxies' velocities \citep{cohen02}.  The primary (right-hand,
``A'') cluster contains significant cold and cooling gas in its core
(a ``cooling flow''). Such gas is seen to be quite robust in simulated
major cluster mergers \citep{gomez02,poole08}. Even in cases where the
cooling flow is finally disrupted by the encounter, \cite{gomez02}
find a delay of $1-2$ GYr between the initial core encounter and the
collapse of the cooling flow. The existence of a strong cooling flow,
therefore, does not argue against a major merger in this case.  More
detailed simulations could shed further light on this interesting
system.

\subsection{Broad Implications}

Since calibrating SZ observable - mass relationships is vital to
understand the implications of ongoing SZE surveys, it is important to
understand the mechanism by which a substantial portion of the ICM can
be heated so dramatically, and how this energy is distributed through
the ICM over time. Observations of cold fronts in other clusters
\citep[e.g.][]{v01} have shown that energy transport processes in the
ICM are substantially inhibited, perhaps by magnetic fields. It is
distinctly possible, then, that once heated by shocks, very hot phases
would persist.  

We have estimated the magnitude of the bias in an arcminute-resolution
Compton $y$ measurement that is introduced by a hot gas phase by
convolving the two gaussian components of the SZE model in
Table~\ref{tbl:szmodel} with a $1'$ FWHM Gaussian beam, typical of SZ
survey telescopes such as ACT \citep{actref,actclus} or SPT
\citep{sptref,sptcat}. Compared with the bulk emission component, also
convolved with a $1'$ beam, the small-scale features are a 10\%
effect. While relatively modest this is a systematic bias in the
Compton $y$ parameter which, if not properly accounted for, would
result in a $20\%$ overestimate in distances (underestimate in $H_0$)
derived from a comparison of the SZE and X-ray data which did not
allow for the presence of the hot gas component. To assess the impact
on the {\it scatter} in $M-y$, a larger sample of high-resolution SZE
measurements is needed. A full calculation would also need to take
into consideration effects such as detection apertures and the spatial
filtering due to imaging algorithms, some of which would increase the
importance of the effect and some of which would decrease its
importance.


This is the one of a very few clusters that has been observed at
sub-arcminute resolution in the SZE \citep[see also][]{nord09}, so it
is possible that many clusters exhibit similar behavior.  Such events,
if their enhancement of the SZE brightness is transient, could also
bias surveys towards detecting kinematically disturbed systems near
the survey detection limits.  The astrophysics that has been revealed
by high resolution X-ray observations, and is beginning to be revealed
by high resolution SZE data, is interesting in its own right.  The SZE
observations require large-aperture millimeter telescopes which have
henceforth been lacking, but with both large single dishes and ALMA
coming online, exciting observations will be forthcoming.  There is
substantial room for improvement: since the observations we report
here the GBT surface has improved from $320 \micron$ RMS to $250
\micron$ RMS, which will yield more than a factor of $1.5$ improvement
in sensitivity. The array used in these observations, while state of
the art, has not yet achieved sky photon noise limited performance;
further progress is being made in this direction. Considering these
facts, and that the results presented here were acquired in a short
period of allocated telescope time (8h), this new high-resolution
probe of the ICM has a bright future.




The National Radio Astronomy Observatory is a facility of the National
Science Foundation operated under cooperative agreement by Associated
Universities, Inc. We thank Eiichiro Komatsu for providing the NOBA SZ
map; Marusa Bradac for providing her total mass map; Masao Sako, Ming
Sun, Maxim Markevitch, Tony Mroczkowski and Erik Reese for helpful
discussions; and Rachel Rosen and an anonymous referee for comments on
the manuscript.


\bibliography{rxj1347mustangApr10}

\end{document}